\documentclass[10pt]{article}

\usepackage[utf8]{inputenc}

\usepackage{amsmath, amsthm, amssymb}
\usepackage{a4wide}
\usepackage{algorithm,algorithmic}
\usepackage{graphicx}
\usepackage{multirow}
\usepackage{url}
\usepackage{arydshln}
\usepackage{subfig}
\usepackage{algorithm,algorithmic}

\begin{document}

\title{Some Comments on the Stochastic Eulerian Tour Problem}
\author{Dennis Weyland \\ Università della Svizzera italiana, Lugano, Switzerland \\ Department of Economics and Management, University of Brescia, Italy \\ \url{dennisweyland@gmail.com}}

\maketitle

\section{Introduction}

The \textsc{Stochastic Eulerian Tour Problem} was introduced in 2008 \cite{mohan2008stochastic} as a stochastic variant of the well-known \textsc{Eulerian Tour Problem}. In a follow-up paper the same authors investigated some heuristics for solving the \textsc{Stochastic Eulerian Tour Problem} \cite{mohan2010heuristics}. After a thorough study of these two publications a few issues emerged. In this short research commentary we would like to discuss these issues. For this purpose, we will first introduce the \textsc{Stochastic Eulerian Tour Problem} using the original formulation of \cite{mohan2008stochastic}. Afterwards we will discuss the following issues more in detail:

\begin{itemize}
 \item The formal definition of the \textsc{Stochastic Eulerian Tour Problem} seems quite cumbersome. In fact, there are many components which are totally meaningless and which distract from the actual problem.
 \item The \textsc{Stochastic Eulerian Tour Problem} has been introduced as a stochastic variant of the well-known \textsc{Eulerian Tour Problem}. This is rather misleading, since the non-stochastic variant of the \textsc{Stochastic Eulerian Tour Problem} is completely different from the \textsc{Eulerian Tour Problem}.
 \item The non-stochastic variant of the \textsc{Stochastic Eulerian Tour Problem} can be easily seen to be a generalization of the famous \textsc{Traveling Salesman Problem}. Hardness and inapproximability results are therefore passed over from the \textsc{Traveling Salesman Problem} to the non-stochastic variant of the \textsc{Stochastic Eulerian Tour Problem} and finally to the \textsc{Stochastic Eulerian Tour Problem}.
 \item In \cite{mohan2008stochastic} it has been stated that the \textsc{Stochastic Eulerian Tour Problem} is an NP-hard stochastic variant of the polynomially solvable \textsc{Eulerian Tour Problem}. This is quite an interesting claim, but based on the previous two issues it has to be rejected. The \textsc{Stochastic Eulerian Tour Problem} is NP-hard but it is neither the stochastic variant of the \textsc{Eulerian Tour Problem} nor the stochastic variant of a polynomially solvable problem.
\end{itemize}

\section{The Stochastic Eulerian Tour Problem}

According to \cite{mohan2008stochastic} the \textsc{Stochastic Eulerian Tour Problem} is defined in the following way. We are given an undirected Eulerian graph $G = (V,E)$, a subset $R \subseteq E$ of the edges with cardinality $n$ and a function $d: E \rightarrow \mathbb{R}$ representing distances on the edges. Additionally, one of the vertices is the depot, where the Eulerian tour is supposed to start and end. In \cite{mohan2008stochastic} this node is duplicated as $v_0$, which then serves as the depot and is connected to the original depot by two edges of length $0$. Furthermore, each edge of the set $R$ has associated a probability which indicates the likelihood that this edge actually requires to be served. Finally, it is assumed that if the number of edges that require to be served in a particular realization of the stochastic data is $k$, then every set of $k$ edges from $R$ is equally likely.

The optimization goal is to compute an a priori Eulerian tour for the graph $G$, such that the expected costs of the a posteriori tour with respect to the given probabilities is minimized. From a given a priori Eulerian tour, the a posteriori tour is derived in the following way. The tour starts at the depot, visits the edges that require to be served in the order and orientation given by the a priori Eulerian tour, and finishes the tour at the depot. For movements between the edges that require to be served, the shortest paths with respect to the given distances are chosen. The cost of such an a posteriori tour is the sum of the traveled distances.

\section{Comments on the Problem Definition}

There are a few issues with the definition of the \textsc{Stochastic Eulerian Tour Problem} which we will discuss in this section. First of all, let us have a look at the a priori solutions. These are Eulerian tours on the given graph $G$. But the only properties of such an a priori solution that are of interested, are the order and orientation of the edges from the set $R$. In fact, the rest of the graph is just used to compute the distances between any of the edges in $R$. After updating the distances between any pair of vertices that belong to edges in $R$ to their shortest path distances within the given graph $G$, we could safely remove the edges which do not belong to the set $R$, that means we could remove all the edges in the set $E \setminus R$. Additionally, we could safely remove all the vertices which do not belong to any of the edges in $R$, except the depot of course. The resulting graph still captures the basic essentials of the original problem. It no longer has to be Eulerian and the a priori solutions are no longer Eulerian tours, but they simply specify an order and an orientation of the remaining edges. It is not clear at all, why the problem was originally defined in such a cumbersome way which even distracts from the important features.

Apart from that, it was originally assumed that if the number of edges that require service in a particular realization of the stochastic data is $k$, then every set of $k$ edges from $R$ is equally likely. This assumption seems really weird, as it immediately implies that the probabilities that edges require service are the same for all edges, which is a highly objectionable restriction of this problem. It might be interesting to investigate this special case, but the problem definition should be more general. As it is done for the presence of vertices for the \textsc{Probabilistic Traveling Salesman Problem}, the presence of edges for the \textsc{Stochastic Eulerian Tour Problem} could be modeled as independent stochastic events. There is no need for the additional assumption given in \cite{mohan2008stochastic}.

On top of these two issues, it is possible to perform another simplification. It definitely makes sense for practical applications to have a depot where the tour starts and ends. But in our simplified definition of this problem such an explicit definition of a depot is no longer necessary, since it can be simulated by an edge of negligible length which always requires service, that means which requires service with a probability of $1$.

All in all, the definition of the \textsc{Stochastic Eulerian Tour Problem} can be simplified to the following definition, which still captures all the basic essentials. We are given an undirected and complete (no longer necessarily Eulerian) graph $G = (V, E)$ and a function $d: E \rightarrow \mathbb{R}$ which represents the distances on the edges. Additionally, we are given a set of edges $R$, such that each vertex from $V$ occurs in exactly one of the edges of $R$, and a function $p: R \rightarrow [0,1]$ which represents the probabilities that the edges require service. Now the objective is simply to find an order and orientation of the edges of $R$, which is the a priori solution, such that the expected costs of the a posteriori solution is minimized. For a particular realization of the stochastic events, the a posteriori solution is derived from the a priori solution by visiting the edges that require service in the order and orientation given by the a priori solution while skipping all the other edges. The cost for such an a posteriori tour is the sum of the traveled distances.

We believe that this definition is much more simple and straightforward than the original definition. We no longer need a depot in this definition. We also do not need the weird assumption about the probabilities. And most important, we do not require the given graph to be Eulerian and the a priori solutions to be Eulerian tours. At the same time it still captures all the basic essentials of the original definition.

\section{The Relation to the Eulerian Tour Problem}

The problem has been named \textsc{Stochastic Eulerian Tour Problem} since it was claimed to be a stochastic variant of the \textsc{Eulerian Tour Problem}. But the main issue here is that the non-stochastic variant of the \textsc{Stochastic Eulerian Tour Problem} is not the \textsc{Eulerian Tour Problem}. This is apparent even for the original definition of the \textsc{Stochastic Eulerian Tour Problem}. Fixing all the probabilities to $1$, we arrive at a problem, where we have to compute a tour of minimum cost, starting and finishing at a depot and visiting certain edges in a specific order and orientation. This is definitely not the \textsc{Eulerian Tour Problem} and as we will see in the next section, this problem differs significantly from the \textsc{Eulerian Tour Problem}. Therefore, it seems very unfortunate that the problem was named the \textsc{Stochastic Eulerian Tour Problem}.

\section{Hardness and Inapproximability of the Stochastic \\ Eulerian Tour Problem}

In \cite{mohan2008stochastic} it was shown that the \textsc{Stochastic Eulerian Tour Problem} is NP-hard by a reduction from the \textsc{Probabilistic Traveling Salesman Problem}. It was not too difficult to prove this result, but with our simplified definition of the problem we can easily show that the \textsc{Stochastic Eulerian Tour Problem} is a generalization of the famous \textsc{Traveling Salesman Problem}. This does not only imply NP-hardness, but it also allows us to transfer inapproximability results from the \textsc{Traveling Salesman Problem} to the \textsc{Stochastic Eulerian Tour Problem}.

For a given instance of the \textsc{Traveling Salesman Problem}, we just have to replace each vertex by two vertices which are sufficiently close to each other and an edge connecting them. These new edges are those which actually require service and we additionally assign a probability of $1$ to them. It is easy to see that there is a straightforward bijection between solutions for this new instance of the \textsc{Stochastic Eulerian Tour Problem} and solutions for the original instance of the \textsc{Traveling Salesman Problem}. Moreover, the costs between such two solutions differ only by a negligible amount. We think that this relation is quite obvious and we do not think that in addition a formal proof is required at this point.

Based on these findings, we have to reject a claim that has been made in \cite{mohan2008stochastic}, namely that the \textsc{Stochastic Eulerian Tour Problem} is an NP-hard stochastic variant of the polynomially solvable \textsc{Eulerian Tour Problem}. Well, the \textsc{Stochastic Eulerian Tour Problem} is NP-hard, but it is neither the stochastic variant of the \textsc{Eulerian Tour Problem} nor the stochastic variant of a polynomially solvable problem. The latter statement follows from the fact that our reduction from the \textsc{Traveling Salesman Problem} was in fact to the non-stochastic variant of the \textsc{Stochastic Eulerian Tour Problem}, since all the probabilities were fixed at $1$.

\section{Discussion and Conclusions}

As we have seen, the definition of the \textsc{Stochastic Eulerian Tour Problem} was in fact quite cumbersome and could be reduced to a much simpler formulation which still captures the basic essentials of this problem. Additionally, we could show that the non-stochastic variant of the \textsc{Stochastic Eulerian Tour Problem} is not the \textsc{Eulerian Tour Problem} but some sort of generalization of the \textsc{Traveling Salesman Problem}. Therefore, hardness and inapproximability results are passed over from the \textsc{Traveling Salesman Problem} to the non-stochastic variant of the \textsc{Stochastic Eulerian Tour Problem} and finally to the \textsc{Stochastic Eulerian Tour Problem}. Based on this result we also have to reject the claim that the \textsc{Stochastic Eulerian Tour Problem} is an NP-hard stochastic variant of a polynomially solvable problem.

\subsection*{Acknowledgments}
This research has been supported by the \emph{Swiss National Science Foundation} as part of the \emph{Early Postdoc.Mobility} grant 152293.

\bibliographystyle{plain}
\bibliography{paper}

\end{document}